\documentclass[11pt,a4paper,onecolumn]{article}
\usepackage{graphicx}
\usepackage{amsmath,amssymb}
\usepackage[colorlinks,linkcolor=blue,citecolor=blue,urlcolor=blue,anchorcolor=blue]{hyperref}
\usepackage{geometry}
\title{Three-dimensional nonparaxial accelerating beams from the transverse Whittaker integral}

\author{Yiqi Zhang$^{1,*}$, Milivoj R. Beli\'c$^{2,**}$, Huaibin Zheng$^{1}$, \\
Haixia Chen$^{1}$, Changbiao Li$^{1}$, Zhiguo Wang$^{1}$, Yanpeng Zhang$^{1,***}$ \\
$^1$\textit{Key Laboratory for Physical Electronics and Devices of} \\ 
\textit{the Ministry of Education \&} \\
\textit{Shaanxi Key Lab of Information Photonic Technique,}\\
\textit{Xi'an Jiaotong University, Xi'an 710049, China}\\
$^2$\textit{Science Program, Texas A\&M University at Qatar,} \\
\textit{P.O. Box 23874 Doha, Qatar }\\
$^*${E-mail: zhangyiqi@mail.xjtu.edu.cn}\\
$^{**}${E-mail: milivoj.belic@qatar.tamu.edu}\\
$^{***}${E-mail: ypzhang@mail.xjtu.edu.cn}}

\date{\today}

\begin{document}

\maketitle

\begin{abstract}
  We investigate three-dimensional nonparaxial linear accelerating beams arising from the transverse
  Whittaker integral. They include different Mathieu, Weber, and Fresnel beams, among other.
  These beams accelerate along a semicircular trajectory, with almost invariant nondiffracting shapes.
  The transverse patterns of accelerating beams are determined by their angular spectra,
  which are constructed from the Mathieu functions, Weber functions, and Fresnel integrals.
  Our results not only enrich the understanding of multidimensional nonparaxial accelerating beams,
  but also display their real applicative potential --
  owing to the usefulness of Mathieu and Weber functions, and Fresnel integrals in describing
  a wealth of wave phenomena in nature.
\end{abstract} 

\section{Introduction}
Accelerating beams have attracted a lot of attention in photonics research community in the last decade
\cite{siviloglou_ol_2007,siviloglou_prl_2007,bandres_oe_2007,ellenbogen_np_2009,chong_np_2010,efremidis_ol_2010,eichelkraut_ol_2010,zhang_ol_2013,zhang_epl_2013,zhang_oe_2014}.
Being solutions of the linear Schr\"odinger equation -- which is equivalent to the
paraxial wave equation in linear dielectric materials --
Airy wave packets from quantum mechanics have made a debut in optics \cite{berry_ajp_1979}.
Paraxial accelerating beams propagating in free space or in linear dielectric media
are related to the Airy or Bessel functions \cite{durnin_josaa_1987,durnin_prl_1987}.
Very recently, it was demonstrated that Fresnel diffraction patterns can also exhibit accelerating properties \cite{zhang_epl_2013}.
Interest in these beams stems from the fact that they exhibit self-acceleration, self-healing, and non-diffraction over many
Rayleigh lengths \cite{siviloglou_ol_2007,siviloglou_prl_2007,broky_oe_2008, ellenbogen_np_2009}.
Following initial reports, accelerating beams by now have been discovered in Kerr media \cite{hu_ol_2010,kaminer_prl_2011,zhang_ol_2013},
Bose-Einstein condensates \cite{efremidis_pra_2013},
on the surface of a gold metal film \cite{minovich_prl_2011}, on the surface of silver \cite{li_prl_2011,li_nano_2011}, in
atomic vapors with electromagnetically induced transparency \cite{zhuang_ol_2012b}, in
chiral media \cite{zhuang_ol_2012a}, photonic crystals \cite{kaminer_oe_2013}, and elsewhere.
However, the transverse acceleration of paraxial accelerating beams is restricted to small angles.
An urge has been created for extending the theory and experiment of accelerating beams beyond paraxial range.

In response, other kinds of linear accelerating beams, called the nonparaxial accelerating beams, have been created quickly
and have attracted a lot of attention recently.
These beams, for example Mathieu and Weber beams, are found by solving the Helmholtz wave equation, and can be
quite involved -- especially in the multidimensional cases \cite{aleahmad_prl_2012,zhang_prl_2012,kaminer_prl_2012,bandres_njp_2013}.
Since the solutions of the three-dimensional (3D) Helmholtz equation can be
represented by a reduced form of the Whittaker integral \cite{whittaker_book},
the 3D accelerating beams with different angular spectra can be investigated using this representation.
If the angular spectrum can be written in terms of the spherical harmonics $Y_{l,m}$,
a 3D accelerating spherical field with a semicircular trajectory can be obtained from such a spectrum \cite{alonso_ol_2012}.
Very recently, 3D accelerating beams coming from rotationally symmetric waves \cite{li_pre_1998},
such as parabolic, oblate and prolate spheroidal fields, have been reported in \cite{bandres_oe_2013}.
In light of nonparaxial accelerating beams ability to bend sharply ($\sim 90^\circ$) or even make U-turns,
they are likely to find applications in micro-particle manipulations.
Thus, nonparaxial accelerating beams deserve further scrutiny.

In this Letter,
we study the 3D nonparaxial accelerating beams by considering different angular spectra introduced into the transverse Whittaker integral.
Because Mathieu and Weber functions figure among the exact solutions of the Helmholtz equation,
it is meaningful to construct the corresponding angular spectra by using the Mathieu and Weber functions directly.
In addition, as the Fresnel integrals are used to describe diffracted waves,
it makes sense to employ them to produce nondiffracting nonparaxial Fresnel accelerating beams.
Thus, in the following we also use the Fresnel integrals to compose an appropriate angular spectrum in the Whittaker integral.
Even though Mathieu, Weber, and Fresnel waves are not fully rotationally symmetric, they still propagate along part-circular trajectories
and represent interesting linear wave systems in physics.
Therefore, here we will investigate novel 3D Mathieu, Weber, and Fresnel nonparaxial accelerating beams together, from the same standpoint.
We should stress that although we use Mathieu and Weber functions to construct angular spectra, the Mathieu and Weber beams reported here
are different from the Mathieu and Weber beams reported elsewhere. This is obvious from the fact that these beams accelerate along
elliptic and parabolic trajectories, whereas our beams accelerate along circular trajectories.


\section{Theoretical Model}
\label{model}

\begin{figure}[htbp]
\centering
  \includegraphics[width=0.3\columnwidth]{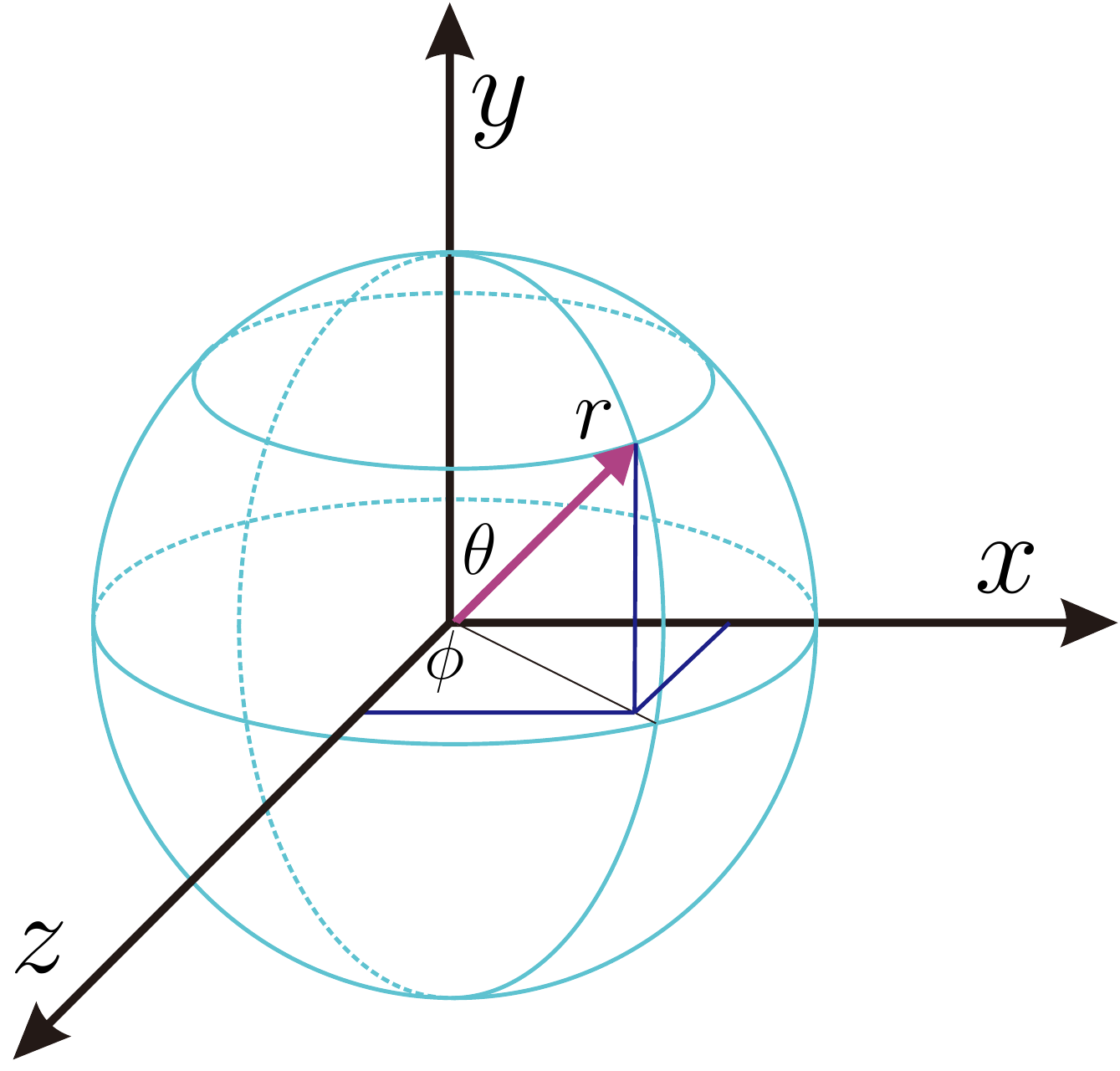}
  \caption{(Color online) Cartesian coordinate ($x,y,z$) and spherical coordinate
($r,\theta,\phi$) with $r$ the radial distance, $\theta$ the polar angle, and $\phi$ the azimuthal angle. }
 \label{axes3d}
\end{figure}

In Cartesian or spherical coordinates as shown in Fig. \ref{axes3d},
the transverse Whittaker integral \cite{whittaker_book,bandres_oe_2013} can be written as
\begin{align}\label{whittaker}
  \psi ({\bf r}) & =\iint \sin\theta d\theta d\phi A(\theta,\phi) \notag \\
  &\times \exp\left[ik(x\sin\theta\sin\phi+y\cos\theta+z\sin\theta\cos\phi)\right],
\end{align}
where $A(\theta,\phi)$ is the angular spectrum function of the wave $\psi$, with $\theta\in [0,\pi]$ and $\phi\in[-\pi/2,\pi/2]$.
Whittaker integral originates from diffraction theory of electromagnetic waves and passes under different names
in different fields, for example as the Debye-Wolf diffraction formula in the imaging theory \cite{wolf_1959}.
Arbitrary spectra are allowed, but we will consider the spectrum functions that can be separated in the variables,
$A(\theta,\phi)=g(\theta)\exp(im\phi)$,
where $g(\theta)$ is a complex function and $m$ is an integer.
Such a form implies rotational symmetry in the problem.
In this case, Eq. (\ref{whittaker}) can be rewritten as
\begin{align}\label{whittaker2}
  \psi ({\bf r}) & = \iint \sin\theta d\theta d\phi [g(\theta)\exp(im\phi)] \notag \\
  &\times \exp\left[ik(x\sin\theta\sin\phi+y\cos\theta+z\sin\theta\cos\phi)\right],
\end{align} which is still quite general.
If we restrict ourselves to the value of the colatitude angle $\theta=\pi/2$,
Eq. (\ref{whittaker2}) will reduce to the two-dimensional case \cite{alonso_ol_2012}.
Utilizing Eq. (\ref{whittaker2}), we can display the transverse intensity distributions
of shape-invariant beams in an arbitrary plane $z=\rm const$
and the accelerating trajectory in an arbitrary plane $y=\rm const$.

Although at this point $g(\theta)$ is an arbitrary function, it is natural to
assume that it is associated with the solutions of the 2D Hemholtz equation.
The procedure is that we first specify solutions of the 2D Helmholtz equation in different coordinates,
and then construct an appropriate angular spectrum function $g(\theta)$.
The corresponding Hemholtz wave eqution is of the form:
\begin{align}\label{hze}
  \left( \frac{\partial^2}{\partial z^2} +\frac{\partial^2}{\partial x^2}+k^2  \right) f(z,x)=0,
\end{align}
the solutions of which include many functions -- the exponential plane wave functions, Bessel functions, Mathieu functions, and Weber
functions that can be expressed in the
Cartesian, circular cylindrical, elliptic cylindrical, and parabolic cylindrical coordinates
\cite{gutierrez-vega_ol_2000,bandres_ol_2004, abramowitz_book}, respectively.
Here $k^2$ is the eignevalue of the corresponding Laplace operator's boundary value problem
(connected with the wavenumber of the corresponding solutions).
In the following sections, we will investigate the three cases for which the angular spectra
are given in forms of the Mathieu and Weber functions, and the Fresnel integrals, respectively.

\section{Mathieu beams}
\label{mathieu}

In the elliptic cylindrical coordinates
$z=h\cosh\xi\cos\eta$ and $x=h\sinh\xi\sin\eta$, with $\xi\in[0,+\infty)$ and $\eta\in[0,2\pi)$,
the solutions of the Helmholtz equation are the Mathieu functions. By utilizing variable separation,
that is, by writing the solution of Eq.
(\ref{hze}) as $f(\xi,\eta)=R(\xi)\Phi(\eta)$, Hemholtz equation separates into
two ordinary differential equations \cite{gutierrez-vega_ol_2000,abramowitz_book}
\begin{subequations}
\begin{align}
\label{mathieu_eq1}\frac{\partial^2 R(\xi)}{\partial \xi^2} - (a-2q\cosh2\xi)R(\xi)=&0,
\end{align}
\begin{align}
\label{mathieu_eq2}\frac{\partial^2 \Phi(\eta)}{\partial \eta^2} + (a-2q\cos2\eta)\Phi(\eta)=&0.
\end{align}
\end{subequations}
where $a$ is the separation constant,
$q=k^2h^2/4$ is a parameter related to the ellipticity of the coordinate system,
$h$ is the interfocal separation,
and $k=2\pi/\lambda$ is the wave number ($\lambda$ being the wavelength in the medium).
The solutions of Eqs. (\ref{mathieu_eq1}) and (\ref{mathieu_eq2}) are the radial and angular Mathieu functions, respectively.
A number of such solutions is known \cite{abramowitz_book}.

\begin{figure}[htbp]
  \centering
  \includegraphics[width=0.6\columnwidth]{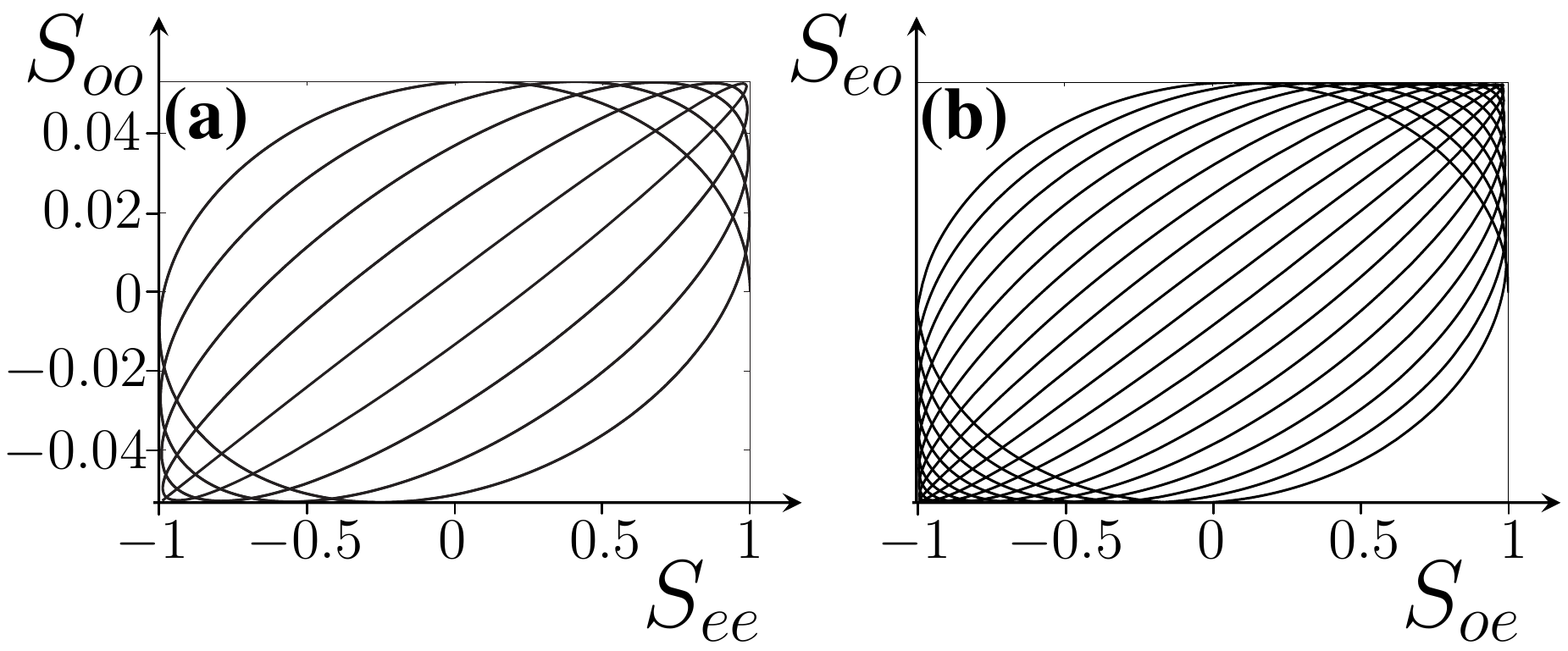}
  \caption{(a) Angular spectrum function $g_1(\theta)$. (b) Angular spectrum function $g_2(\theta)$.
  The order of the Mathieu functions is 10, and $q=4$.}
  \label{mathieu_spectrum}
\end{figure}

We utilize the angular Mathieu functions to construct the corresponding spectral function $g$.
Generally, there are four categories of the angular Mathieu solutions of Eq. (\ref{mathieu_eq2}), denoted as
$S_{ee}, ~S_{eo}, ~S_{oe}$, and $S_{oo} $ for the even-even, even-odd, odd-even, and odd-odd cases, respectively;
they have been discussed in detail in \cite{abramowitz_book}. They also carry an order $m$, which can be quite high
in the cases considered here; the solutions then look quite similar and for convenience we suppress the order index $m$.
Thus, the term $g(\theta)$ in the angular spectrum of Eq. (\ref{whittaker2}) may be written in the following way:
\begin{align}
\label{mathieu_g1}  g_1(\theta) & = S_{ee}(\theta) +i S_{oo}(\theta), \\
\label{mathieu_g2}  g_2(\theta) & = S_{oe}(\theta) +i S_{eo}(\theta),
\end{align}
and in this manner one may introduce the novel Mathieu accelerating beams, by utilizing the transverse Whittaker integral.
This is in difference to the Mathieu nonparaxial beams introduced elsewhere \cite{zhang_prl_2012}, where combinations of the radial
and angular Mathieu functions have been utilized.
Before going to details, we show the angular spectrum functions $g_1$ and $g_2$ in Figs.
\ref{mathieu_spectrum}(a) and  \ref{mathieu_spectrum}(b).
The order of the involved Mathieu functions is 10, with $q=4$.

\begin{figure}[htbp]
  \centering
  \includegraphics[width=0.6\columnwidth]{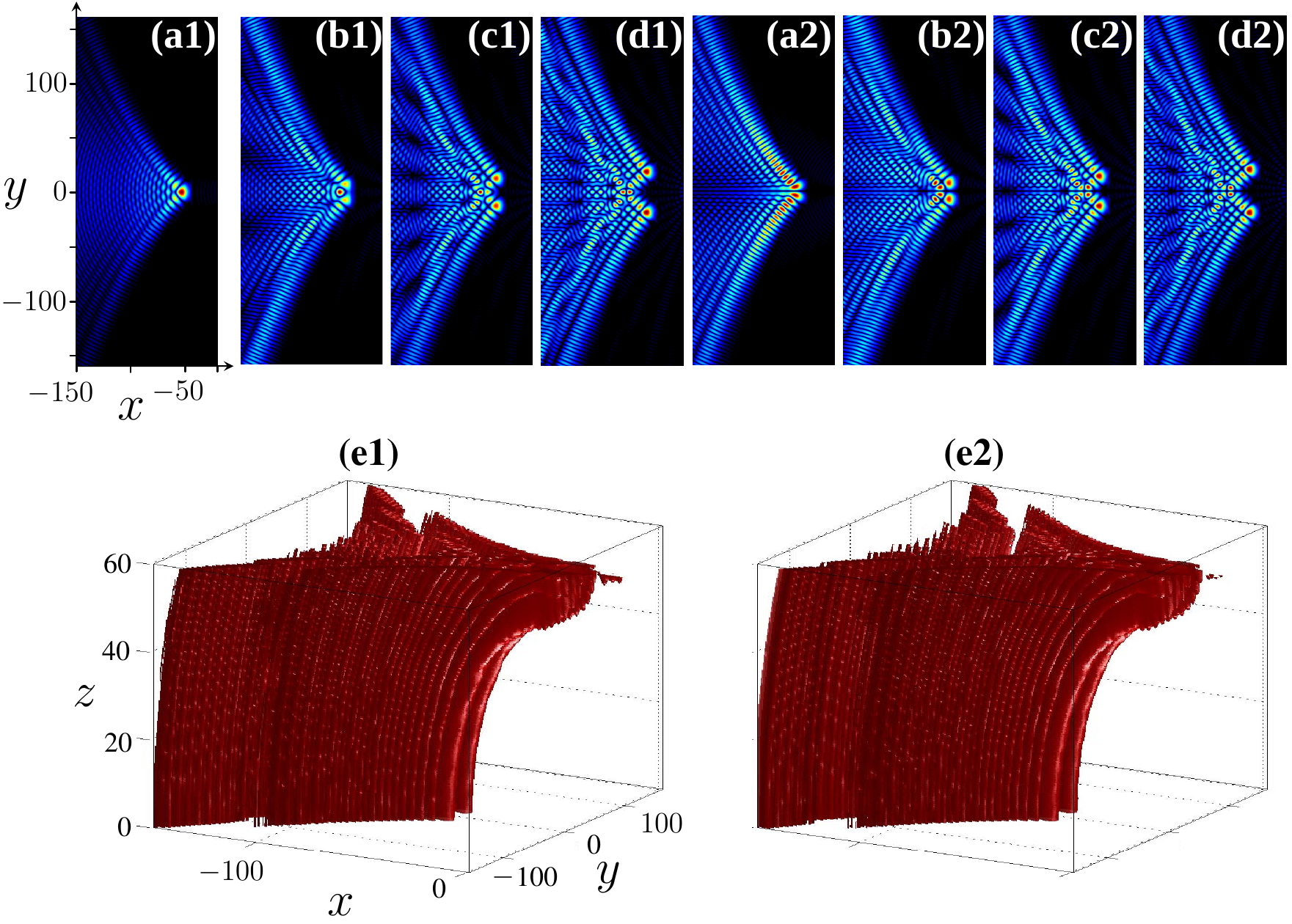}
  \caption{(Color online) Cross-section in the $z=0$ plane of the amplitude $|\psi(\bf r)|$ of the shape-invariant Mathieu
  beams with $q=4$ and $m=50$. The 1st, 4th, 7th, and 10th order Mathieu functions are displayed from left to right.
  (a1)-(d1) The case of $g_1$. (a2)-(d2) The case of $g_2$.
  (e1) and (e2) 3D plots of the propagating 10th Mathieu beam, corresponding to $g_1$ and $g_2$, respectively.}
  \label{mathieu_beams}
\end{figure}

Based on the spectral functions $g_1$ and $g_2$, we display the transverse $|\psi(\bf r)|$ distributions of the shape-invariant Mathieu beams
at $z=0$ in Figs. \ref{mathieu_beams}(a1)-\ref{mathieu_beams}(d1)
and Figs. \ref{mathieu_beams}(a2)-\ref{mathieu_beams}(d2), respectively.
It is seen that the amplitude distributions $|\psi(\bf r)|$ of the beams in the $z=0$ plane are symmetric about $y=0$,
and that they become quite complex as the order is increased.
Different from the $g_1$ case, the amplitude $|\psi(\bf r)|$ at the $y=0$ plane is always the smallest for the $g_2$ case.
To observe the accelerating properties as the beams propagate, we choose the 10th order Mathieu functions as an example;
numerical simulations are shown in Figs. \ref{mathieu_beams}(e1) and \ref{mathieu_beams}(e2),
which correspond to the cases presented in Figs. \ref{mathieu_beams}(d1) and \ref{mathieu_beams}(d2), respectively.
It is seen that the beams are almost invariant along the $z$ direction,
slowly expanding and accelerating along a quarter-circular trajectory. Note that the beams mirror-extend
below the $z=0$ plane, so that the beams in the $y=const$ plane accelerate along a semicircular trajectory.
This comes from the following argument.

As indicated in Sec. \ref{model}, we set the polar angle $\phi$ in the interval
$[-\pi/2,~\pi/2]$; that means the beam propagates in the positive $z$ direction.
If $\phi\in[\pi/2,~3\pi/2]$, the beam will propagate in the negative $z$ direction,
and it will also accelerate along a quarter-circular trajectory, which is symmetric about the plane $z=0$
to the positive quarter-circular trajectory.
When both the positively and negatively propagating beams are considered together,
one observes a beam accelerating along a semicircular trajectory.
For convenience, we only consider the case of $\phi\in[-\pi/2,~\pi/2]$ throughout the Letter.
From the 3D plot of the acceleration process,
we see that the beams exhibit a nonparaxial accelerating property during propagation -- as they should.

\section{Weber beams}
\label{weber}

The solutions of the Helmholtz equations in the parabolic cylindrical coordinates are the Weber functions.
The relation between the Cartesian and parabolic cylindrical coordinates is
$z+ix=(\xi+i\eta)^2/2$, with $\xi\in[0,+\infty)$ and $\eta\in(-\infty,+\infty)$.
When Eq. (\ref{hze}) is expressed in the parabolic cylindrical coordinates, and the method of separation of variables applied,
the Hemholtz equation is transformed into a system of equations \cite{zhang_prl_2012,bandres_ol_2004,abramowitz_book}
\begin{subequations}
\begin{align}
\label{weber_eq1}\frac{\partial^2 R(\xi)}{\partial \xi^2} + \left( k^2\xi^2-2ka \right) R(\xi)=&0,
\end{align}
\begin{align}
\label{weber_eq2}\frac{\partial^2 \Phi(\eta)}{\partial \eta^2} + \left( k^2\eta^2+2ka \right) \Phi(\eta)=&0,
\end{align}
\end{subequations}
in which $2ka$ is the separation constant.
If one sets $\sqrt{2k}\xi=u$ and $\sqrt{2k}\eta=v$,
Eqs. (\ref{weber_eq1}) and (\ref{weber_eq2}) transform into the two standard Weber's differential equations
\begin{subequations}
\begin{align}
\label{weber_eq3}\frac{\partial^2 R(u)}{\partial u^2} + \left( \frac{u^2}{4} - a \right) R(u)= &0,
\end{align}
\begin{align}
\label{weber_eq4}\frac{\partial^2 \Phi(v)}{\partial v^2} + \left( \frac{v^2}{4} + a \right) \Phi(v)= &0.
\end{align}
\end{subequations}
Therefore, the solutions of Eqs. (\ref{weber_eq1}) and (\ref{weber_eq2}) are the Weber functions with different eigenvalues.
If we denote the even and odd solutions of Eq. (\ref{weber_eq2}) as $P_e$ and $P_o$, we can define the angular spectrum function $g(\theta)$ as
\begin{align}\label{weber_g}
  g(\theta)=P_e(\theta,a) + iP_o(\theta,a),
\end{align}
in which
\begin{subequations}
\label{eq10}
\begin{align}\label{Peo}
  P_{e,o}(\theta,a)=& \sum^\infty_{n=0} c_n \frac{\theta^n}{n!},
\end{align}
and the coefficients $c_n$ satisfy the following recurrence relation:
\begin{align}\label{Peoc}
  c_{n+2} = & ac_n - \frac{n(n-1)c_{n-2}}{4}.
\end{align}
\end{subequations}
To obtain $P_e$ or $P_o$, one sets the first two $c_n$ coefficients to $c_0=1, c_1=0$ or to $c_0=0, c_1=1$, respectfully \cite{bandres_ol_2004}.
This choice for the angular spectrum functions is again different from the choice made in \cite{zhang_prl_2012}, in which the combinations of solutions of
both Eqs. (\ref{weber_eq1}) and (\ref{weber_eq2}) are used to construct the full angular spectrum of Weber accelerating beams.
These Weber beams accelerate along parabolic trajectories and are different from the ones introduced here.

\begin{figure}[htbp]
\centering
  \includegraphics[width=0.6\columnwidth]{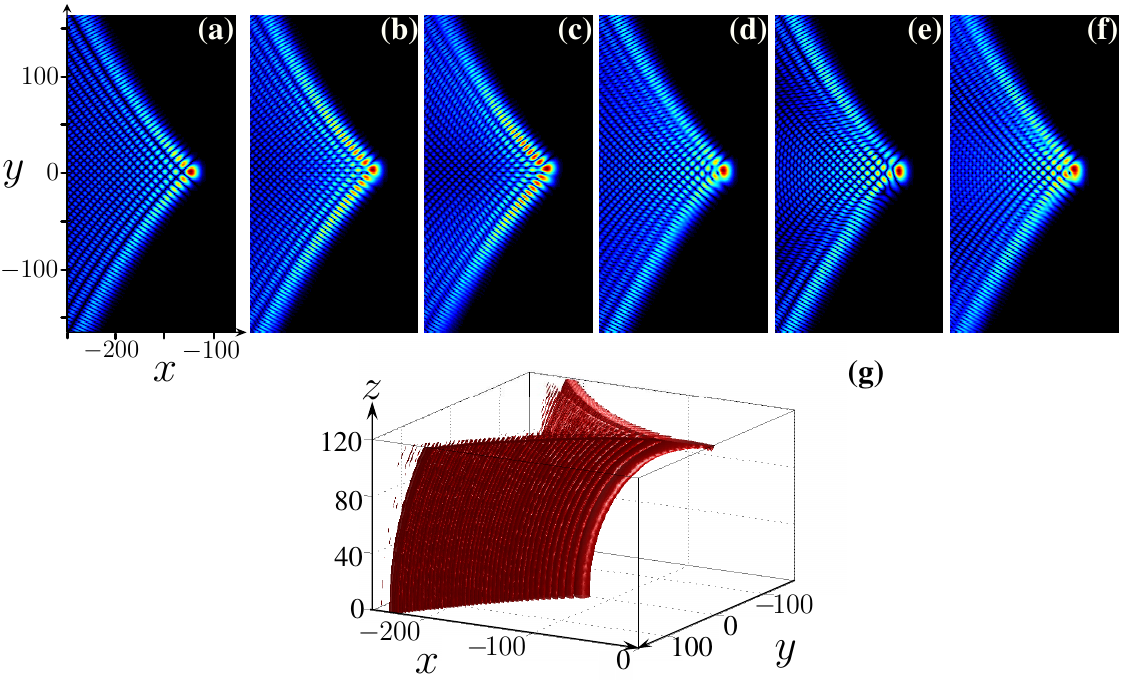}
  \caption{(Color online) (a)-(f) Wave amplitude $|\psi(\bf r)|$ distribution at $z=0$ of the shape-invariant Weber beams,
  for $m=120$ and $a=1,~7,~10,~13,~16$ and $19$, respectively.
  (g) 3D plot of the Weber beam with $a=1$.}
  \label{weber_AB}
\end{figure}

To display our Weber beams, we repeat the same procedure as for the Mathieu beams; the only difference is that now
there are not two $g$ functions but one, expressed in terms of the even and odd Weber eigenfunctions.
The amplitude $|\psi(\bf r)|$ distribution of Weber beams at $z=0$ for $m=120$ and different $a$, according to Eqs. (\ref{weber_g}) and (\ref{eq10}),
is shown in Figs. \ref{weber_AB}(a)-\ref{weber_AB}(f).
It is seen that the $|\psi(\bf r)|$ distribution changes slightly with $a$ increasing, especially around the point $(x=-m,~y=0)$.
Regardless of what the $|\psi(\bf r)|$ distribution at the $z=0$ plane is, the beam still accelerates along a quarter-circular trajectory.
As an example, Fig. \ref{weber_AB}(g) depicts the accelerating process of the beam with $a=1$, which corresponds to Fig. \ref{weber_AB}(a).
It is clear that the shape of the beam is nearly invariant as it propagates,
and the accelerating trajectory is a quarter of the full circle in the plane $y=0$
(the bending angle is approximately $90^\circ$). So, the acceleration is nonparaxial.

\section{Fresnel integrals}
\label{fresnel}

\begin{figure}[htbp]
\centering
  \includegraphics[width=0.6\columnwidth]{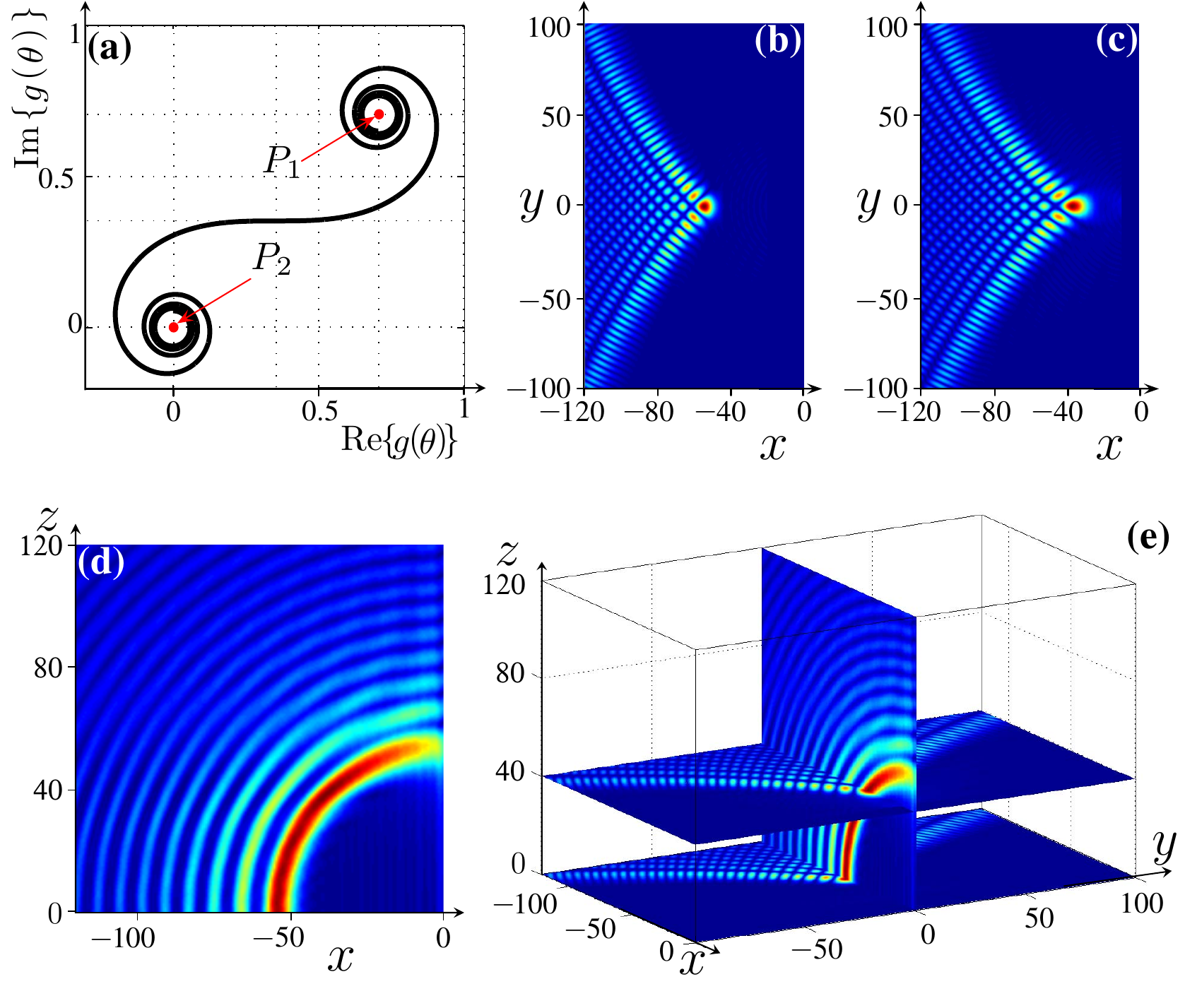}
  \caption{(Color online) (a) Spectrum function $g(\theta)$ of Fresnel integrals.
  (b) Transverse intensity distribution of the beam with $m=50$ at $z=0$.
  (c) Same as (b) but at $z=40$.
  (d) Acceleration along the circular trajectory of the beam at the cross-section at the plane $y=0$.
  (e) The panels (c), (d) and (e) put together.}
  \label{fresnel_fig}
\end{figure}

In optics, Fresnel integrals are commonly used to describe Fresnel diffraction from a straight edge or a rectangular aperture.
The Fresnel cosine and sine integrals are defined as \cite{born_book}
\begin{subequations}
\begin{align}
\label{fresnel_cosine}\mathcal{C}(t)=&\int^t_0 \cos\left(\frac{\pi}{2}\tau^2\right)d\tau,
\end{align}
\begin{align}
\label{fresnel_sine}\mathcal{S}(t)=&\int^t_0 \sin\left(\frac{\pi}{2}\tau^2\right)d\tau,
\end{align}
\end{subequations}
or in the complex form:
\begin{equation}
\mathcal{F}(t)=\mathcal{C}(t)+i\mathcal{S}(t).
\end{equation}
Thus, we define the new Fresnel nonparaxial accelerating wave by introducing
the spectrum function $g(\theta)$ as
\begin{align}
\label{fresnel_g}g(\theta)=\left(\mathcal{C}(\theta)+\frac{1}{2}\right)+i\left(\mathcal{S}(\theta)+\frac{1}{2}\right).
\end{align}
These Fresnel waves are different from the ones introduced in \cite{zhang_epl_2013}, which propagate paraxially
along parabolic trajectories.
The spectral function $g(\theta)$ is shown in Fig. \ref{fresnel_fig}(a), which is the Cornu spiral \cite{born_book}; the
two branches of the spiral approach the points $P_1$ and $P_2$, with the coordinates $(1/\sqrt{2},1/\sqrt{2})$ and $(0,0)$.

According to Eq. (\ref{fresnel_g}) and the Whittaker integral, we can now present the Fresnel nonparaxial accelerating beams.
We display the intensity distributions of the accelerating beam at $z=0$ plane, $z=40$
plane, and $y=0$ plane in Figs. \ref{fresnel_fig}(b), \ref{fresnel_fig}(c), and \ref{fresnel_fig}(d), respectively.
Concerning the intensity at $z=0$ plane, it is similar to the ones shown in Figs. \ref{mathieu_beams}(a1) and \ref{weber_AB}(a),
even though the angular spectra are quite different.
In comparison with Figs. \ref{fresnel_fig}(b) and \ref{fresnel_fig}(c), we find the beam accelerating --
all the lobes move rightward along the positive $x$ direction.
In Fig. \ref{fresnel_fig}(d) we can follow the acceleration of the main lobe and other higher-order lobes in the $y=0$ plane.
It is evident that the lobes accelerate along a quarter-circular trajectory, with a bending angle $\sim90^\circ$,
which demonstrates that the acceleration is nonparaxial. As before, the beams in Fig. \ref{fresnel_fig}(d) extend to
the region $z\le0$, so that the acceleration actually proceeds along semicircular trajectories.
In order to see the accelerating process more clearly,
we put Figs. \ref{fresnel_fig}(b), \ref{fresnel_fig}(c) and \ref{fresnel_fig}(d) together in Fig. \ref{fresnel_fig}(e).

\section{Conclusion}
\label{conclusion}

Based on the transverse Whittaker integral, we have investigated the novel 3D nonparaxial accelerating beams,
by constructing the angular spectra using Mathieu functions, Weber functions, and Fresnel integrals, respectively.
In the transverse plane, the intensity distributions of the accelerating beams are related to the angular spectra.
In the longitudinal direction,
the beams accelerate along a quarter-circular trajectory, with the bending angle approximately $90^\circ$.
Even though the angular Mathieu functions, Weber functions, and Fresnel integrals used here are not strictly rotationally symmetric,
investigation on such 3D nonparaxial accelerating beams is still meaningful and deserves appropriate attention.
Since Mathieu and Weber functions are the solutions of the Helmholtz equation, and the
Fresnel integrals are used to describe optical diffraction,
which are all related to common phenomena in real world, we hope to have revealed the importance of such beams for
potential applications.

\section*{Acknowledgement}
This work was supported by the 973 Program (2012CB921804),
CPSF (2012M521773),
and the Qatar National Research Fund NPRP 6-021-1-005 project.
The projects
NSFC (61308015, 11104214, 61108017, 11104216, 61205112),
KSTIT of Shaanxi Province (2014KCT-10),
NSFC of Shaanxi Province (2014JZ020, 2014JQ8341),
XJTUIT (cxtd2014003),
KLP of Shaanxi Province (2013SZS04-Z02),
and FRFCU (xjj2013089, xjj2014099, xjj2014119)
are also acknowledged.

\bibliographystyle{unsrt}
\bibliography{refs}

\end{document}